\documentclass[]{spie} 
\usepackage[]{graphicx}
\usepackage{amsmath}
\usepackage{amssymb}
\usepackage{hyperref}

\title{Wavefront sensing from the image domain with the Oxford-SWIFT integral field spectrograph}

\author{Benjamin Pope\supit{a}, Niranjan Thatte\supit{a}, Rick Burruss\supit{b}, Matthias Tecza\supit{a}, Fraser Clarke\supit{a}, Garret Cotter\supit{a}
\skiplinehalf
\supit{a}Department of Astrophysics, Denys Wilkinson Building, University of Oxford, OX1 3RH, UK \\
\supit{b}Jet Propulsion Laboratory, Pasadena, California, USA
}

\authorinfo{Corresponding Author: Benjamin Pope, benjamin.pope@astro.ox.ac.uk}
 
\begin{document} 
  \maketitle 

\begin{abstract}
The limits for adaptive-optics-assisted and space-based astronomical imaging at high contrast and high resolution are typically determined by residual phase errors due to non-common-path aberrations not sensed by the wavefront sensor. These impose quasi-static speckles on the image, which are difficult to calibrate as they vary in time and with telescope orientation. Typical approaches require phase diversity of some sort\cite{2012SPIE.8447E..21K}, which requires many iterations and is accordingly time-consuming.  This is especially true of integral field spectrographs, where use of standard phase diversity based techniques is additionally complicated by the presence of the image slicer / integral field unit.

We present the first application of the kernel phase based `asymmetric pupil Fourier wavefront sensing' scheme to ground-based AO-corrected integral field spectroscopy, whereby an asymmetric pupil mask and a single image are sufficient to map aberrations up to high order, including non-common-path error. This method is closely connected with kernel phase interferometry, already applied to space-based and AO-assisted imaging, in which a phase transfer matrix formalism partitions focal plane Fourier phases into a kernel space which is self-calibrating with respect to pupil aberrations, and a row space which can be used to determine those aberrations via a matrix pseudo-inverse. This requires two key conditions be satisfied: the first, that phase errors are $<$~1 radian in magnitude. These conditions are typically satisfied for space-based telescopes such as the HST, or AO-corrected ground-based telescopes in the near-infrared. The second requirement is that the telescope pupil is not centro-symmetric; this can be achieved simply by placing an asymmetric mask in the optical path. The row phase reconstruction then provides a phase map which can be applied directly to a deformable mirror as a static offset. While in our approach we have iteratively applied corrections, we have deliberately damped correction steps, and in principle this can be done in a single step. 

We push toward internally diffraction-limited performance with the Oxford-SWIFT integral field spectrograph coupled with the PALM-3000 extreme AO system on the Palomar 200-inch telescope. This represents the first observation in which the PALM3000 + SWIFT internal point-spread-function has closely approached the Airy pattern. While this can only be used on SWIFT with an internal stimulus source, as at short wavelengths the uncorrected atmospheric wavefront errors are still $>$~1 radian, this nevertheless demonstrates the feasibility of detecting non-common-path errors with this method as an active optics paradigm for near-infrared, AO-corrected instruments on Palomar such as PHARO or Project 1640 (P1640), or other instruments such as VLT-SPHERE or the Gemini Planet Imager (GPI). We note that this is a particularly promising approach for correcting integral field spectrographs, as the diversity of many narrowband images provides strong constraints on the wavefront error estimate, and the average of reconstructions from many narrow bands can be used to improve overall reconstruction quality. 
\end{abstract}



\section{INTRODUCTION}
\label{intro}  

A central problem in astronomy is to achieve the highest possible resolution and contrast in optical imaging and imaging spectroscopy. The primary limitation on image quality is imposed by the turbulence of the atmosphere, or `seeing', which degrades images on scales finer than $\sim$ several arcseconds. The resultant point spread function (PSF) consists of individual speckles on timescales shorter than a characteristic seeing timescale $t_0$, which average out to an approximately Gaussian radial distribution on longer timescales.

This can be ameliorated with adaptive optics (AO) systems\cite{1953PASP...65..229B}, which use a wavefront sensor (WFS) to measure wavefront aberrations as seen on a nearby guide star, and corrects for these with a deformable mirror (DM) in a closed feedback loop \cite{2012ARA&A..50..305D}. These guide stars may be a natural guide star, i.e. an unresolved astronomical source nearby in the sky, or a laser guide star may be used, whereby a patch of the sodium layer is caused to fluoresce with a laser. If we measure the image quality by the Strehl ratio, defined as the ratio of the peak measured intensity to the theoretical diffraction-limited intensity for an object of the same intrinsic flux, such AO systems are typically able to reach a Strehl ratio of $\sim 30-60 \%$ in the near-infrared. 

A further improvement is gained by extreme AO systems where both a low-order and a high-order deformable mirror (LODM and HODM respectively) are used to correct aberrations up to very high order. Such systems have been shown to reach a Strehl ratio of $\sim 80-90 \%$ at near-IR wavelengths. 

The main limitation on the performance of such extreme adaptive optics systems is from non-common-path (NCP) errors, which lie in the differences of the optical paths between the DM and WFS, and between the DM and science detector. These are inaccessible to measurement with the WFS and typically vary on long timescales, causing quasi-static speckles that are difficult to calibrate. These long-lived speckles are not easily distinguished from real astrophysical objects, and as such limit the sensitivity achievable in imaging at high spatial resolution and contrast, such as is needed for studies of stellar and planetary-mass faint companions.

In such a case, it is necessary to derive information about the NCP errors from the science instrument itself. It is not, in general, possible to uniquely determine the wavefront giving rise to a particular PSF, for the reason that the autocorrelation of a function is not uniquely invertible. We therefore normally need more information to sense these aberrations. The typical approach to calibrating these speckles is to obtain images both in and out of focus, and use this phase diversity to solve the otherwise-underdetermined problem\cite{1994ApOpt..33.6533K,2004OptL...29.2707C,2012OptL...37.4808S}.

Recently, a new approach to wavefront sensing from the image domain has been proposed \cite{2013PASP..125..422M}, using a linear phase transfer approximation to reconstruct a pupil phase map using only a single image taken with the science camera. This approximation was first introduced in the context of imaging \cite{2010ApJ...724..464M}, proposing `kernel phases' as a generalization of closure phases for refining high-angular-resolution imaging with a well-corrected aperture. These kernel phases are self-calibrating observables, consisting of the kernel space of the phase transfer matrix. These linear combinations of $uv$-plane phases are then robust with respect to small wavefront errors. Crucially, this approximation requires that residual phase aberrations are smaller than 1 radian, which renders the technique applicable only to instruments with no need for atmospheric correction (e.g. space telescopes) or with extreme AO. Kernel phase interferometry has already been used to anchor models of binary systems \cite{2013ApJ...767..110P} and presents a promising new direction in space-based and AO-corrected imaging and imaging spectroscopy studies. 

The wavefront sensing paradigm developed here  \cite{2013PASP..125..422M} arises from the same formalism, whereby a pseudoinverse can be found to the same phase transfer matrix. Phases from the complementary subspace to the kernel (hereafter the kernel complement) can then be mapped uniquely back onto the pupil aberrations with no need for phase diversity in exposures. An important requirement for this is that the pupil itself must be asymmetric with respect to inversion through its centre; as such, it has been called the asymmetric pupil Fourier wavefront sensor. This has previously been experimentally demonstrated using a microelectromechanical (MEMS) deformable mirror \cite{2014MNRAS.440..125P}, but has not previously been attempted on a facility adaptive optics system and science instrument. In the previous case, it was shown to restore a significantly degraded wavefront (Strehl $\lesssim 60\%$) to a state not experimentally distinguishable from the diffraction limit.

\section{Palomar Observatory Experimental Apparatus}
\label{palomar}

In this experiment, we have applied the asymmetric pupil Fourier wavefront sensing approach to correcting the NCP error on the Oxford-SWIFT integral field spectrograph and Palomar PALM-3K extreme adaptive optics system. 

\subsection{PALM-3000 Extreme Adaptive Optics}
\label{p3k}

The Palomar PALM-3000 (hereafter P3K) adaptive optics system \cite{2013ApJ...776..130D} is the successor to the original PALM-241 (P241) system on the Palomar 200-inch telescope, using a Shack-Hartmann WFS and a dual LODM and HODM to achieve extreme AO correction. The key innovation in P3K is the introduction of a HODM with a $66 \times 66$ actuator grid, of which 3388 active actuators span the instrument pupil. The LODM is the 241-active-actuator system that previously comprised P241. The introduction of the HODM extends the outer AO control radius from $4 \lambda/D$ to $32 \lambda/D$, thus enabling very high contrast ($< 10^{-7}$ with P1640\cite{2013ApJ...776..130D}) to be achieved in direct imaging searches for faint companions.

P3K achieves residual wavefront errors (excluding NCP) of $\sim 140$ nm under 1" (median) seeing conditions. We note that at this level of wavefront error, the linear phase transfer approximation holds overall for a 900~nm working wavelength, and permits kernel phase operation on-sky in the absence of NCP, as well as with the calibration source described in Section~\ref{swift}.

\subsection{The Oxford-SWIFT Integral Field Spectrograph}
\label{swift}

The Oxford-SWIFT integral field spectrograph (IFS), hereafter SWIFT, is an image slicing IFS on the Palomar 200-inch telescope in southern California. SWIFT operates in the wavelength range between 650 and 1000 nm in a single exposure, at a spectral resolving power ranging from $R = 3250$ at 650 nm to $R = 4400$ at 1000 nm. It obtains spectra on a grid of  $89 \times 44$ spaxels with three available spatial sampling scales: 235~mas, 80~mas and 16~mas. The last of these Nyquist samples the diffraction limit for the 200-inch telescope for wavelengths longer than 800 nm. It possesses an internal white-light fibre stimulus source for calibration, and it is this source which we have used in the expreriments described in Section~\ref{experiment}. This stimulus can be introduced into the focal plane with subpixel positioning accuracy and is point-like on scales probed by the SWIFT spatial scale, permitting us to directly measure the instrumental PSF.

Even in the absence of atmospheric seeing, using only the P3K stimulus source, SWIFT in general fails to approach the diffraction limit due to internal aberrations of the AO and science instrument. In AO-corrected operation, these will appear as NCP errors which cannot be detected with the WFS and which vary in an unknown way over time. As such, there is no single static calibration that can be used to correct these, and they must be separately determined in each observing run. In the past, the standard approach has been to hand-tune the PSF by trial and error using low-order (astigmatism, coma and defocus) corrections. We note that this error is typically $\gtrsim 1$~radian, and as such typically lies outside the kernel phase regime and an initial attempt must be made to reduce this to within tractable limits. As discussed in Section~\ref{experiment}, this has caused difficulty for the SWIFT experiments, but will typically be achievable more easily on instruments with smaller initial NCP error. This arises primarily from astigmatism on the dichroic splitting off the short wavelength channel to SWIFT.

\subsection{HODM Mask}
\label{sec:mask}

\begin{figure}[h!]
\center
\includegraphics[scale=0.1]{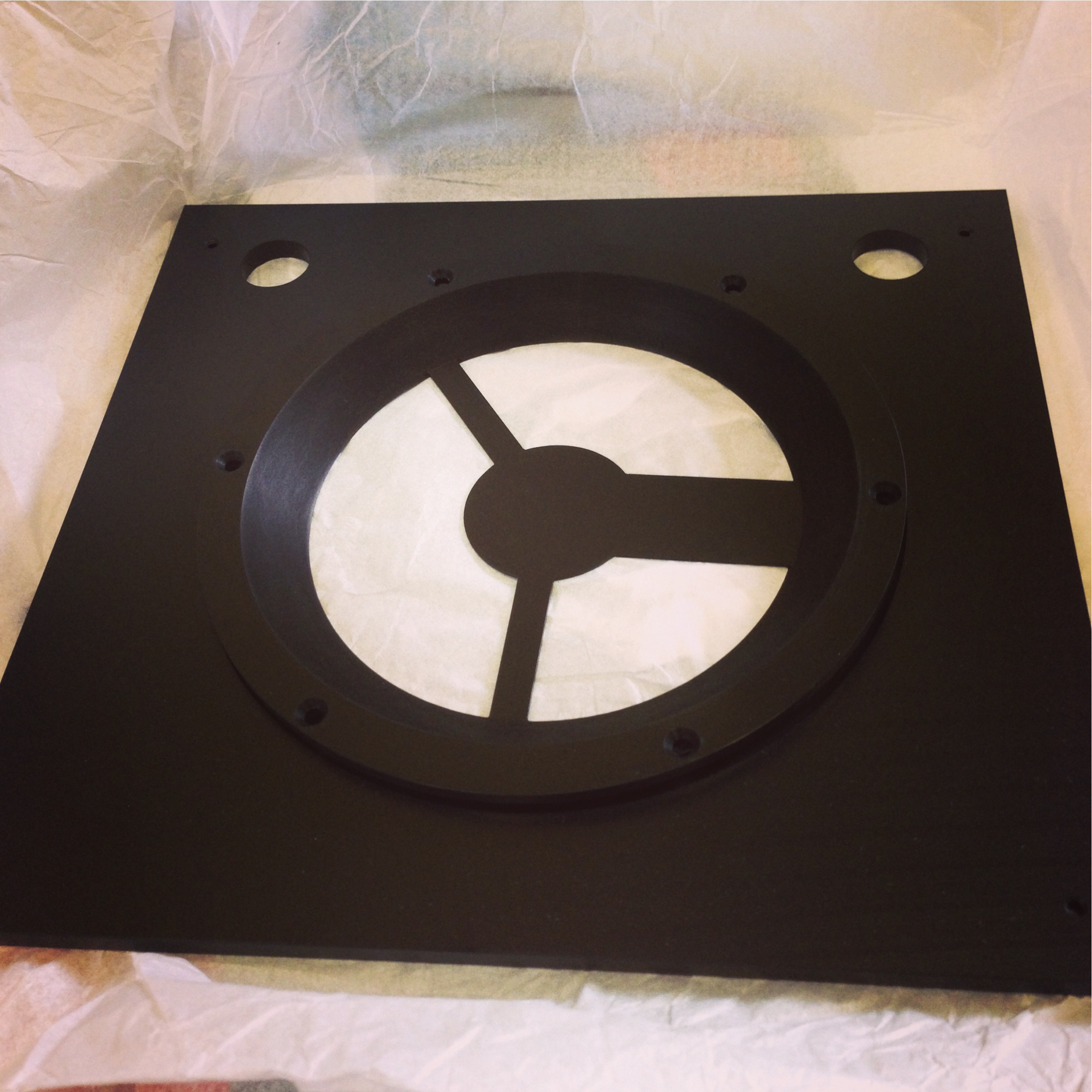}
\caption{HODM Mask before installation.}
\label{maskim}
\end{figure}

In order to provide the required pupil asymmetry, we designed and produced a pair of non-inversion-symmetric anodized aluminium masks for the HODM mount. An example is shown in Figure~\ref{maskim}. The mask designs were simply chosen for mechanical reasons, in particular, requiring three spiders in order to make the mask sufficiently stiff to avoid contact with the HODM at all times. It remains a subject for future work to optimize these masks to allow for the highest throughput while still being sufficiently asymmetric to generate a full basis of kernel complement phases.

\subsection{Software}
\label{software}

All data cubes were first reduced using the standard SWIFT pipeline in IRAF developed by R. Houghton \footnote{\url{http://www-astro.physics.ox.ac.uk/instr/swift/pdfs/SWIFTPipelineManual_v1p43.pdf}}, correcting for flat field illumination, wavelength calibration and bias subtraction.

Wavefront sensing and AO control software was based off the \texttt{pysco} Python Self-Calibrating Observables module \cite{2014MNRAS.440..125P}. This requires a discrete pupil model, which we define a priori from the HODM mask design as noted in Section~\ref{sec:mask}, together with parameters to describe its rotation and parity with respect to the pupil as seen by the SWIFT focal plane. This map is then used to precalculate the transfer matrix and its Moore-Penrose pseudoinverse, which is stored and used for the remainder of the experiment. 100 wavelength channels sampling at 1 \AA/channel are binned to provide 10 nm bandpasses centred on 900 nm, 940 nm and 970 nm. Because the spaxel sampling of 16 mas/spaxel only Nyquist-samples the diffraction limit at wavelengths longer than 800 nm, we restricted ourselves to these longer wavelengths in order to very strictly avoid the effects of any spaxel scale uncertainties.

Wavefront maps are therefore defined first on this grid of pupil samples, and interpolated by the nearest-neighbour method onto the $66 \times 66$ HODM actuator grid as described in Section~\ref{p3k}. This interpolated wavefront map has non-zero values in regions obscured by the mask, which is a potential source of uncorrected error when the mask is removed and the full pupil used to examine the unobscured PSF. This map is sent to the AO control computer and used to generate centroid offsets which add an additional static correction to the wavefront as measured by the Shack-Hartmann spot centroids.

\section{Experiment}
\label{experiment}

We have conducted three experiments on three separate observing runs on the Palomar 200-inch telescope to attempt to correct the NCP error on SWIFT, in October and December 2013, and in April 2014. Of these, on the third occasion we failed to achieve the $< 1$~rad initial NCP error required to begin correcting the wavefront with kernel phase, and as such we do not discuss those results here.

On the October observing run, the HODM electronics had recently failed and it was therefore unavailable for our experiment. This therefore severely limited the accessible aberration modes to those addressed by the LODM, and consequently we did not expect to restore diffraction limited performance beyond the first Airy rings. In this experiment, we used the white light stimulus source to obtain the instrumental PSF, which required the telescope to point to zenith. As the NCP is partially dependent on telescope orientation, it would therefore be a significant improvement in future to reconfigure the stimulus to be accessible at any elevation, and therefore to be able to tune the NCP accurately for observing targets anywhere in the sky.

On the first day of the experiment, we installed the HODM mask and defocused the telescope, and took an exposure with the stimulus source in order to observe the pupil plane. From this we determined that the thick spider of the mask was oriented at $\sim 45^\circ$ to the horizontal. At this point, we set aside the experiment to perform astronomical observations.

Returning to the experiment, we began with the HODM unobscured. We tuned the low-order terms by hand to the point that the PSF appeared from visual inspection to be approaching the kernel phase regime. Next, it was important to determine the inversion parity of the pupil preceding the SWIFT focal plane, as this switches through focus. By applying artificial additional astigmatism and coma using our own centroid offset control code, we were able to determine that there was a single flip involved and adjusted this parameter in the code accordingly, as described in Section~\ref{software}. 

\begin{figure}[h!]
\center
\includegraphics[scale=0.6]{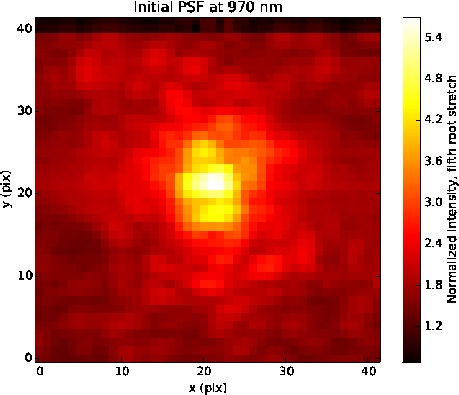}
\caption{10-nm bandwidth image centred on 970 nm, before kernel phase correction.} 
\label{iminit}
\end{figure}

\begin{figure}[h!]
\center
\includegraphics[scale=0.6]{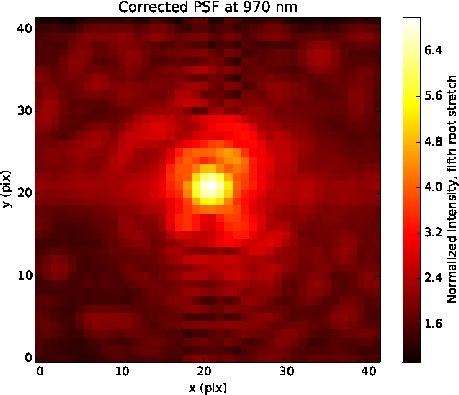}
\caption{10-nm bandwidth image centred on 970 nm, after kernel phase correction.}  
\label{im970}
\end{figure}

After we were satisfied with the open-mask PSF quality (shown in Figure~\ref{iminit}), we installed the HODM mask and took an exposure. Reducing this as discussed in Section~\ref{software}, we then calculated a wavefront map and applied the centroid offsets accordingly. An improvement in the concentration of flux towards the centre of the PSF was immediately noticeable, and we repeated this for four more iterations until it appeared to have converged. We then removed the HODM mask and exposed once more to check the PSF of the unobscured pupil. A 10 nm bandpass centred on 970 nm is showing in Figure~\ref{im970}, showing a bright central spot, a clear first Airy ring and signs of a second. Given that only the LODM was operational, we hypothesised that the remaining NCP error was due to higher order aberrations and paused the experiment until the next observing run.

We returned for a second attempt in December, after the HODM had been repaired. We began in the same way by verifying the instrument and pupil orientation had not changed, and again by tuning the low order aberrations by hand until we were satisfied that the kernel phase approximation would hold. We then repeated the experiment conducted in October, iterating until convergence. The resulting PSF was as compact as previously, and still displayed only one clear Airy ring, even with full high-order wavefront control. We discuss possible explanations for this result in Section~\ref{conclusions}. 

\section{Conclusions and Future Directions}
\label{conclusions}

As Figure~\ref{im970} shows, we have achieved low-order aberration correction with the kernel complement approach, without the need for phase diversity. This is a key advantage of this technique: whereas other methods for focal plane wavefront sensing necessarily depend on a sequence of images taken with phase diversity, which may be in an open or closed loop, in this case, it is possible to determine the total wavefront error including NCP with a single image. This is especially advantageous for instruments like SWIFT with long (180 s) readout times, where it would be prohibitively time-consuming to obtain the diversity required for other techniques. 

While we have not achieved the high-order performance achieved elsewhere\cite{2014MNRAS.440..125P}, we expect that this is due to a combination of technical difficulties which we discuss below. 

The first possibility is that the pupil scale and orientation were insufficiently accurate to permit high-order sensing. Estimation of the orientation of the mask relative to the pupil as seen by SWIFT was, as noted in Section~\ref{experiment}, accurate to only $\sim 5^\circ$. Given that this only corresponds to a few actuators' position error even at the edge of the pupil, it would be surprising if this made a significant contribution to the uncorrected errors. Nevertheless, we expect that in future it will be important to establish this metrology with greater precision. It may alternatively be the case that while we have measured the orientation of the LODM, there may be additional terms between the LODM and HODM for which we have not corrected. Future experiments must therefore account for the geometry along the full optical path.

A second possibility is that the spatial sampling of 16 mas/spaxel was too close to the Nyquist limit, and that improving this by drizzling \cite{2002PASP..114..144F} a set of images together, it will be possible to improve the sampling rate (up to a maximum factor of 2, with four half-pixel steps) and improve the kernel phase performance. We have attempted simulations of such drizzled images (in prep.), and have found a small improvement in the accuracy of wavefronts reconstructed using this method, but not enough to wholly fix the problems discussed in Section~\ref{experiment}. 

The very narrow field of view may also be a problem, in that by aggressively windowing the image to much less than the total $32 \lambda/D$ interferometric field of view, this is equivalent to a convolution in the Fourier domain corrupting phase information. This will in particular introduce $\pm \pi$ phases at the high spatial frequency cutoff, bleeding in from regions where the $u,v$ amplitude is formally zero. This may therefore be an issue that will be solved when moving to instruments with a significantly larger field of view.

In conclusion, we have shown low-order NCP error calibration with an integral field spectrograph, with no need for phase diversity. While it is not immediately clear from these experiments why we have been limited to correcting low-order aberration, the algorithm has been previously demonstrated to sense wavefront errors up to high order \cite{2014MNRAS.440..125P}. We therefore see no reason why we should not expect this technique to be successful provided these technical issues are addressed, and we believe this may become a new and important tool in the calibration of present and future instruments, particularly integral field spectrographs.

\acknowledgments   

We would like to acknowledge the helpful advice of Frantz Martinache and Peter Tuthill, Ryan Houghton for the use of the SWIFT IRAF data reduction package, James Lynn for fabricating the HODM mask, and the careful assistance of Kevin Rykoski and Carolyn Heffner at Palomar Observatory.

The Oxford SWIFT integral field spectrograph is directly supported by a Marie Curie Excellence Grant from the European Commission (MEXT-CT-2003-002792, Team Leader: N. Thatte). It is also supported by additional funds from the University of Oxford Physics Department and the John Fell OUP Research Fund. Additional funds to host and support SWIFT were provided by Caltech Optical Observatories. 

This paper is based in part on observations obtained at the Hale Telescope at Palomar Observatory as part of a collaborative agreement between the California Institute of Technology, its divisions Caltech Optical Observatories and the Jet Propulsion Laboratory (operated for NASA), and Cornell University.


\bibliography{ms}  
\bibliographystyle{spiebib} 

\end{document}